\documentclass[twoside]{article}
\usepackage[accepted]{aistats2016}

\usepackage{url}
\usepackage{amsmath}
\usepackage{amsthm}
\usepackage{amssymb}
\usepackage{graphicx}
\usepackage{caption}
\usepackage{subcaption}
\usepackage{multirow}
\usepackage{booktabs}
\usepackage{pifont}

\newcommand{\argmin}{\mathop{\mathrm{arg\,min}}}

\newcommand*\rot{\rotatebox{90}}

\providecommand{\myfloor}[1]{\left \lfloor #1 \right \rfloor }

\begin{document}



\twocolumn[

\aistatstitle{Sparse Density Representations for Simultaneous Inference on
Large Spatial Datasets}

\aistatsauthor{ Taylor B. Arnold}

\aistatsaddress{ Yale University } ]

\begin{abstract}
Large spatial datasets often represent a number of spatial
point processes generated by distinct entities or classes of events.
When crossed with covariates, such as discrete time buckets, this can
quickly result in a data set with millions of individual density estimates.
Applications that require simultaneous access to a substantial subset
of these estimates become resource constrained when densities
are stored in complex and incompatible formats. We present a method for
representing spatial densities along the nodes of sparsely populated trees.
Fast algorithms are provided for performing set operations and queries on
the resulting compact tree structures. The speed and simplicity of
the approach is demonstrated on both real and simulated spatial data.
\end{abstract}

\section{Introduction and motivation}

\begin{figure*}[t!]
    \centering
    \begin{subfigure}[b]{0.32\textwidth}
        \centering
        \includegraphics[height=1.8in]{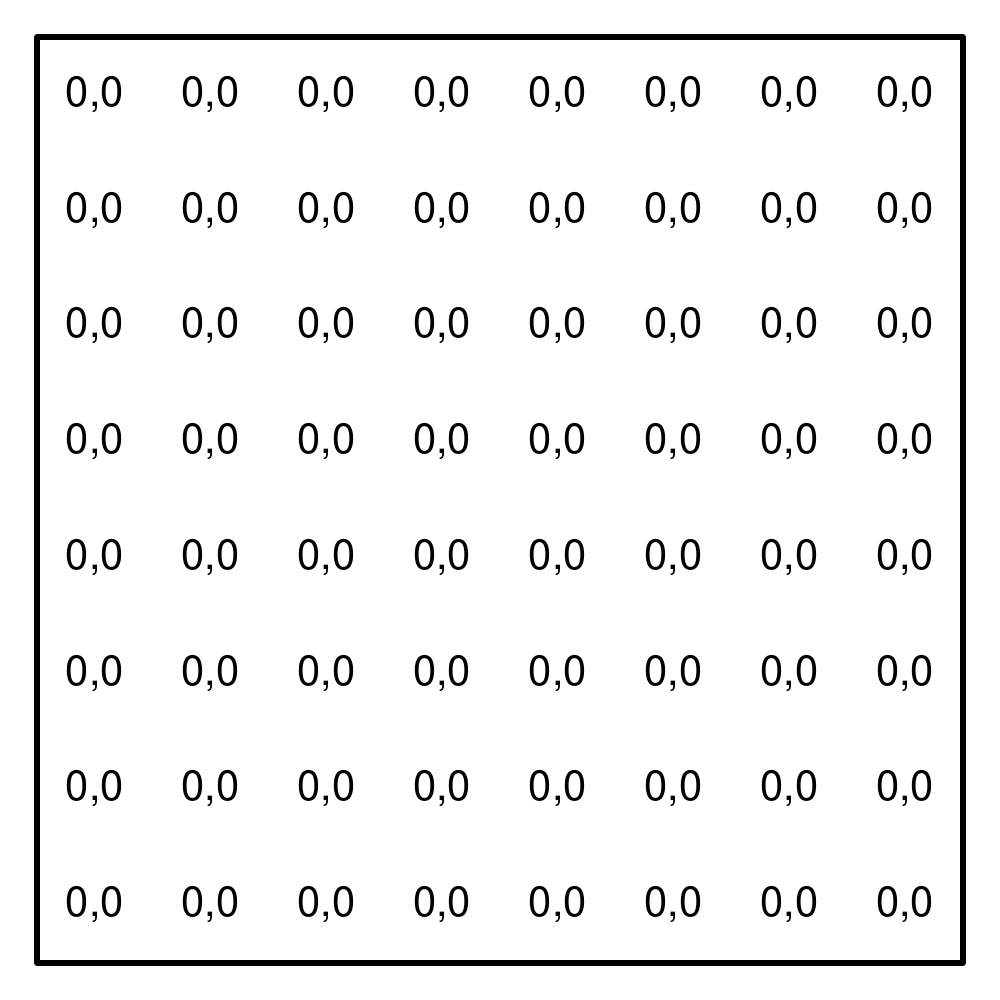}
        \caption{$m^0 = 0$}
    \end{subfigure}%
    ~
    \begin{subfigure}[b]{0.32\textwidth}
        \centering
        \includegraphics[height=1.8in]{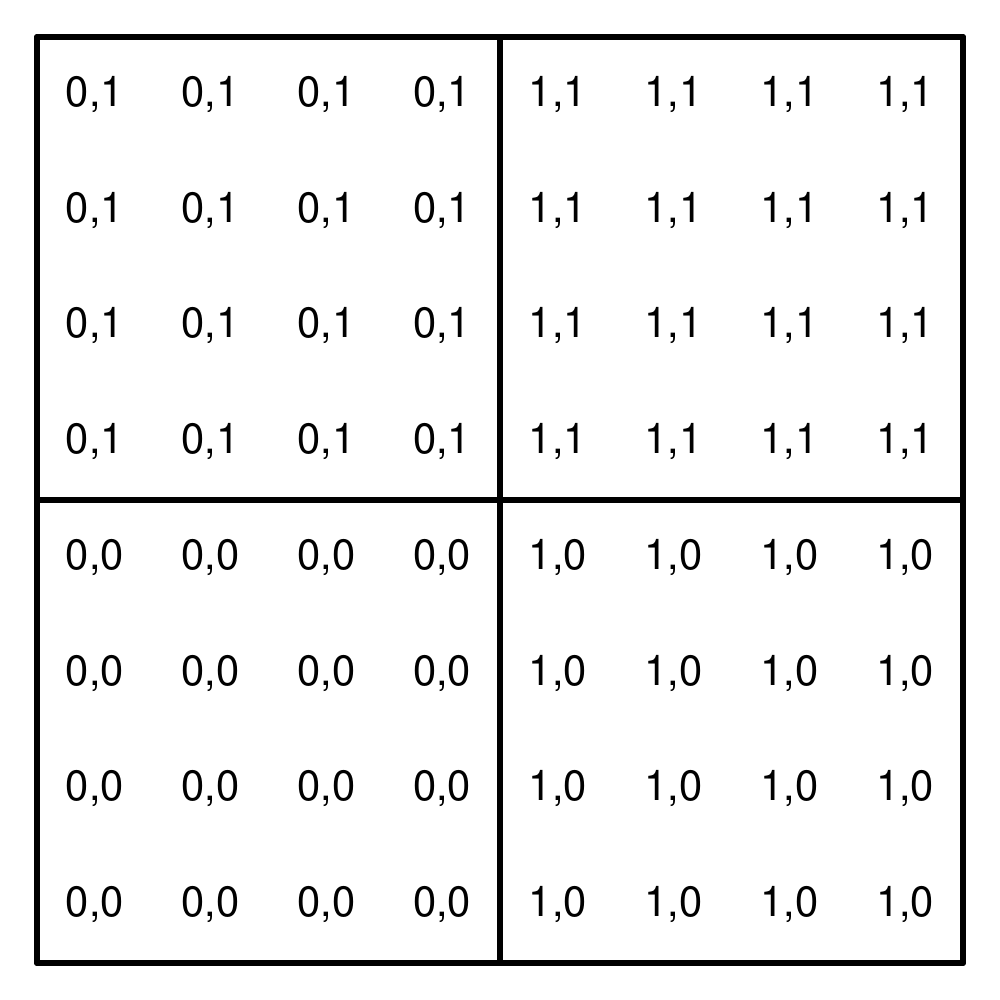}
        \caption{$m^0 = 1$}
    \end{subfigure}%
    ~
    \begin{subfigure}[b]{0.32\textwidth}
        \centering
        \includegraphics[height=1.8in]{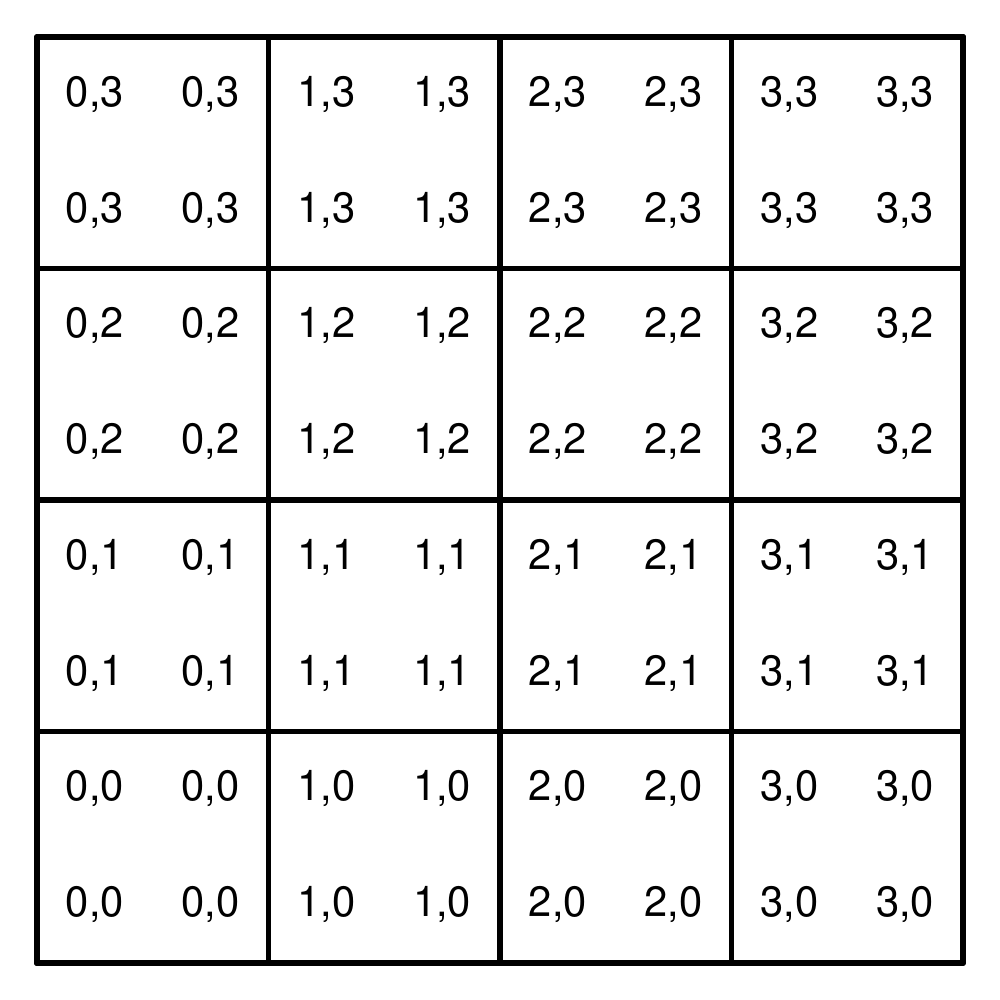}
        \caption{$m^0 = 2$}
    \end{subfigure}%
    \caption{Partitioning of a $2^3$ by $2^3$ grid into quadtree
    based groups. The pair of numbers printed on each grid point
    represent the corrisponding $m^1$ and $m^2$ coordinates. Note that
    there is one final set of partitions for $m^0=3$, where each
    point is in its own tile.}
    \label{tilepart}
\end{figure*}

Many large streaming datasets are generated by systems that continually
record the location of spatially-centric events. Examples include the
address where ambulances are requested through an emergency dispatch
center, location names detected in a query through a web search engine,
and the GPS coordinates recorded by a taxicab's routing software.
When spatial measurements come from a relatively small and discrete set,
they can often be handled just like any other covariate. Spatial data
recorded from a very large set, such as addresses, or as a continuous
measurement of latitude and longitude, may require more specialized
treatment. Predicting the location of the next observation, for example,
is often accomplished by modeling historical data as a spatial point
process and estimating a density function with techniques such as
kernel density estimators or Gaussian mixtures.

With enough data, it may be both possible and desirable to generate
alternative density estimates for different classes of events. When
crossing several variables, densities can quickly become quite
granular; consider, for example,the density of ambulance requests
`related to traffic accidents from 7am-8am on Tuesdays'. An increase
in granularity comes with a corresponding increase in the number
of density estimates that must be stored. On its own, this increase
is not a particular point of concern. The estimates can be stored
in an off-the-shelf database solution, a small subset relevant to
a given application can be queried, and the results used in a
standard fashion.

Computational issues do arise when there is a need to simultaneously
work with a large set of densities. Inverse problems, where an event
is detected in a location and the type of event must be predicted,
present a common examples of such an application. Performing set
operations, such as intersections and unions, to construct even
more complex density estimates is another example. A large set
of non-parametric density estimates, represented by predictions
over a fine-grained lattice, can lead to a substantial amount
of memory consumption. Parametric estimates, perhaps compactly
represented by Gaussian mixtures, may be safely loaded into memory
but eventually cause problems as set operations between densities
yield increasingly complex results. These make inference and visualization
intractable as the scale of the data increases.

As a solution to these problems, we present a method that combines the
benefits of parametric and non-parametric density estimates for large
spatial datasets. It represents densities as a sum of uniform densities
over a small set of differently sized tiles, thus yielding a
sparse representation of the estimated model. At the same time, the space
of all possible tiles is a relatively small and fixed set; this allows
multiple densities to be aggregated, joined, and intersected in a natural
way. By using a geometry motivated by quadtrees we are also able to support
constant time set operations between density estimates.

\section{Prior work}

The estimation of general density estimation has a long history, with
many spatial density routines being a simple application of generic
techniques to the two-dimensional case. Mixture models were studied
as early as 1894 by Karl Pearson \cite{pearson1894contributions}, with
non-parametric techniques, such as kernel density estimation, being
well known by at least the 1950s \cite{parzen1962estimation},
\cite{rosenblatt1956remarks}.
Further techniques have also been developed in the intervening years,
such as smoothed histograms \cite{gu2003penalized},
splines \cite{boneva1971spline,eilers1996flexible}, Poisson regression \cite{eilers1996flexible},
and hierarchical Bayes \cite{villela2015bayesian}.
A variant on mixture models that allow for densities to depend on additional
covariates was proposed by Schellhase and Kauermann \cite{schellhase2012density},
though their work focuses only on one-dimensional models.

Computational issues regarding density estimation have also been
studied in the recent literature. Some attention has focused
on calculating the kernel density estimates given by a large underlying
training dataset \cite{raykar2010fast, zheng2013quality}. As an active
area of computer vision, there as been a particular interest on two-dimensional
problems \cite{cortes2015sparse, yang2003improved, zheng2015error}. These
often utilize some variation of the fast Gauss transform of
Greengard and Strain \cite{greengard1991fast}. To the best of our knowledge,
no prior work has focused on the computational strains of working simultaneously
with a large set of density estimates.

A significant amount of literature and software also exists from
the perspective of manipulating generic spatial data within a database.
These store spatial data in a format optimized for some set of spatial
queries, such as k-nearest neighbors or spatial joins \cite{yeung2007spatial},
and can be made to handle and query fairly large sets of data. Recent developments have
even made spatial queries possible over data distributed across horizontally
scalable networks \cite{nidzwetzki2015distributed}, and integrated for
fast real-time visualizations \cite{lins2013nanocubes}.
The elements in such systems are typically points, lines, or polygons.
Density functions can be encoded into such a system by storing the centroid
of the density modes within mixture models, or saving predictions over a
fine grid for non-parametric models. The problem in using this solution for
our goal is that
operations on the parametric models do not translate into fast queries on
the database, and the size of the grid required to store non-parametric
models quickly grows prohibitively large. Our solution builds a density
estimate that can be queried within a database system, but stored in a
significantly smaller space.

\section{Density estimation}

\subsection{Approach}

For the remainder of this article, we focus on density estimation over a
rectangular grid of points. We assume that there is some observed sample
density $y_i$ over the grid, which generally corresponds to assigning
observations to the nearest grid point, and an initial estimate of the density
given by $z_i$. The latter typically comes from a kernel density estimate,
but may be generated by any appropriate mechanism. Our goal is to calculate
predictions $x_i$ such that the following are all true:
\begin{enumerate}
\item The new estimates are expected to be nearly equally as predictive
for a new set of observed data $y_{new}$ as the estimates $z_i$,
\begin{align}
\mathbb{E} || y_{new} - z || \approx \mathbb{E} || y_{new} - x ||
\end{align}
\item The estimates $x_i$ can be represented as a sparse vector over some
fixed dictionary $\mathcal{D}$, which has a total dimension of size
$\mathcal{O}(n)$, the number of data points in the grid.
\item Queries of the form $\sum_{i\in A} x_i$ can typically be calculated
faster than $\mathcal{O} (|A|)$ for $A$ corresponding to any subset of points
over the original grid.
\end{enumerate}
To satisfy these requirements, we construct an over-complete,
hierarchical dictionary and calculate an estimate of $z$ using an
$\ell_1$-penalized regression model.

\subsection{Quadtree dictionary}

In order to describe our target dictionary, it is best to consider the
case where the grid of points is a square with $2^k$ points on each side.
We otherwise embed the observed grid in the smallest such square and
proceed as usual knowing that we can throw away any empty elements in the
final result.

For any integer $m^0$ between $0$ and $k$, we define the following sets:
\begin{align}
\mathcal{T}^{m^0}_{m1,m2} &= \left\{i : \myfloor{\frac{i \, (\text{mod} \, 2^k)}{2^{m^0}}} = m_1, \myfloor{\frac{i}{2^{k+m^0}}} = m_2\right\} \nonumber \\
\forall &\, m^1,m^2 \in \left\{0, 1, \ldots, 2^{m^0} - 1 \right\}
\end{align}
For a fixed $m^0$, these sets produce a disjoint, equally sized, partition of the
grid of points, with $m^1$ and $m^2$ giving the horizontal and vertical
coordinates of the superimposed grid. We refer to each $\mathcal{T}^{m^0}_{m1,m2}$
as a {\it tile}, as it represents a square subset of the original lattice.
For a visualization of the partitioning scheme over a small grid, see
Figure~\ref{tilepart}. It will be helpful to have a way of referencing the set of all tiles
that share a given $m^0$ parameter, often referred to as a zoom level:
\begin{align}
\mathcal{T}^{m^0} &= \left\{\mathcal{T}^{m^0}_{m1,m2}, \, \text{s.t.} \, m^1,m^2 \in \left\{0, 1, \ldots, 2^{m^0} \right\} \right\}
\end{align}
Combining all of the zoom levels, we can construct the set of all tiles from
which we will define the final dictionary:
\begin{align}
\mathcal{T} &= \bigcup_{m^0=0}^{k} \mathcal{T}^{m^0}
\end{align}
We will assume that there is some fixed order of this set, so that we may
refer unambiguously to the $j$-th tile $\tau_j$ in the set $\mathcal{T}$. The elements
$m^0_j$, $m^1_j$, and $m^2_j$ refer to the corresponding indices of that
$j$-th tile.
The set $\mathcal{T}$ consists of all tiles corresponding to a fixed depth
quadtree over the grid of all points. They mirror the structure
of tiles used in tilemap servers, such as slippy map generated by
OpenStreetMap \cite{sample2010tile}. When the grid is defined over the entire
globe, our tiles directly coincide with the slippy tiles.

Finally, we define the dictionary $D^{\alpha}$ as a sparse matrix in
$\mathbb{R}^{4^k \times d}$, with $d = |\mathcal{T}|$. It is defined
such that:
\begin{align}
D_{i,j}^{(\alpha)} &= \begin{cases}
    4^{-m^0_j \cdot \alpha},& \text{if } x_i \in \tau_j \\
    0,              & \text{otherwise}
\end{cases}
\end{align}
For $\alpha$ equal to $1$, element $D_{i,j}^{(1)}$ encodes the proportion
of the tile $\tau_j$ that is covered by the point $x_i$. The element
$D_{i,j}^{(0)}$ is instead a simple indicator for whether a the point $x_i$
is in tile $\tau_j$. Other values of $\alpha$ provide a continuous scale of
weightings to the tiles that moves between these two extremes, and will
be useful in the penalized estimation routine.
Notice that the dictionary is an over-representation of the space of
all possible estimators, and for $\alpha$ equal to $0$, contains a
copy of the identity matrix as a permuted subset of its columns.

\subsection{Estimation algorithm}

In order to learn a sparse representation of $z$ given our dictionary
$D^\alpha$, we first use an $\ell_1$-penalized estimator. This well-known
technique produces a parsimonious estimator via convex optimization. The
parameter $\alpha$ is set to a number between $0$ and $1$ to control
the degree to which the penalty should be proportional to tile size.
A large parameter will yield a model with many small tiles, whereas a small
parameter gives a smaller model with more large tiles. We find that a value
near $0.5$ typically works well.
\begin{align}
\widehat{b^{1}} &= \argmin_{b} \left\{ || z - D^{(\alpha)}b ||_2^2 + \lambda \cdot ||b||_1 \right\} \label{ell1}
\end{align}
In order to solve Equation~\ref{ell1}, we calculate the regularization
path for a sequence of $\lambda$ values using software that provides
a customized application of coordinate decent \cite{friedman2010regularization}.
Choosing the final tuning parameter can be done by a number of methods; we have
found the one standard deviation rule, originally suggested by Breiman, with 5-fold
cross validation provides good predictability without overfitting the
model \cite{breiman1995better}.

To increase interpretability and reduce prediction bias, we calculate
the non-negative least squares estimator over the support of the
$\ell_1$-penalized estimator. With a large training dataset, this should
be refit on a holdout set. Empirically, we observe good performance even
when refitting on the same training data. Without the $\ell_1$-penalty
the choice of $\alpha$ does not effect the predicted values, so here
we use $D^1$ to facilitate the interpretability and normalization of
the results.
\begin{align*}
\widehat{b^{2}} = \argmin_{b} &\left\{ || z - D^{(1)}b ||_2^2 \quad \text{s.t.} \right. \nonumber \\
& \quad \left.\text{supp}(b) = \text{supp}(\widehat{b^{1}}) \quad \text{and} \quad b_j \geq 0 \right\}
\end{align*}
We then hard threshold the non-negative least squares solution by a
value of $\delta$.
\begin{align}
\widehat{b^{3}}_j &= \begin{cases}
    \widehat{b^{2}_j},& \text{if } \widehat{b^{2}_j} > \delta \cdot ||\widehat{b^{2}}||_1 \\
    0,              & \text{otherwise}
\end{cases}
\end{align}
And finally, the density estimator is normalized to have a sum of $1$:
\begin{align}
\widehat{b} &= \frac{\widehat{b^{3}}}{||\widehat{b^{3}}||_1}
\end{align}
The predicted values, $x$, can be calculated by projecting $\widehat{b}$
by the dictionary $D^{(1)}$.
\begin{align}
x &= D^{(1)} \widehat{b}
\end{align}
Clearly we do not want to save the predicted values explicitly. Otherwise,
we would have simply saved the raw predicted values $z$. We instead show,
in the next subsection, that predictions $x$ can be generated quickly
from the sparse vector $\widehat{b}$.

\begin{figure*}[t!]
    \centering
    \includegraphics[height=3in]{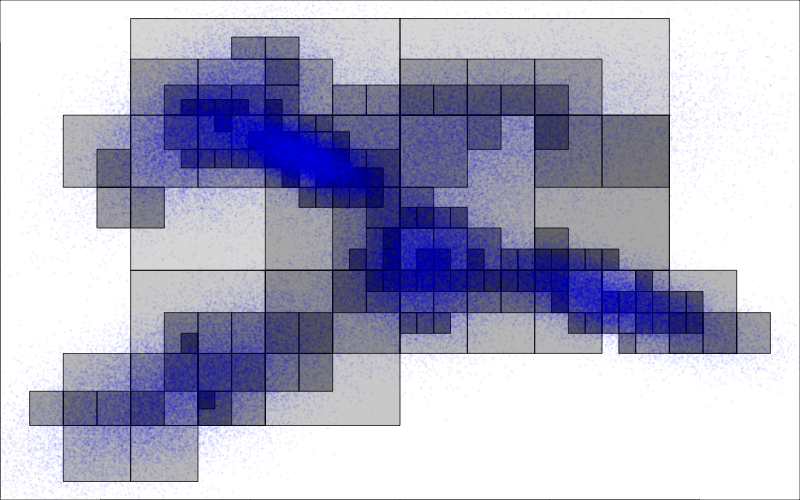}
    \caption{Scatter plot of the training data from the simulated
    Gaussian mixture model superimposed over the predicted densities
    from tile model with grid size equal to $4^7$ and $\alpha$ equal
    to $0.8$.}
    \label{sim1image}
\end{figure*}

\begin{table*}[t!]
\centering
\begin{tabular}{c|ccccccccc|c}
  \hline
num.  & \multicolumn{9}{|c|}{$\alpha$ parameter} & \multirow{2}{*}{$\bar{y}$} \\
tiles & 0 & 0.1 & 0.25 & 0.33 & 0.5 & 0.66 & 0.75 & 0.9 & 1 & \\
  \hline
  $4^3$ & 0.662 & 0.662 & 0.566 & 0.566 & 0.714 & 0.714 & 0.454 & 0.714 & 0.809 & 0.137 \\
  $4^4$ & 0.177 & 0.266 & 0.266 & 0.266 & 0.200 & 0.148 & 0.147 & 0.808 & 0.808 & 0.134 \\
  $4^5$ & 0.212 & 0.211 & 0.210 & 0.202 & 0.189 & 0.195 & 0.198 & 0.914 & 0.974 & 0.178 \\
  $4^6$ & 0.267 & 0.258 & 0.241 & 0.239 & 0.242 & 0.250 & 0.252 & 0.263 & 0.979 & 0.217 \\
  $4^7$ & 0.320 & 0.322 & 0.308 & 0.307 & 0.312 & 0.315 & 0.319 & 0.338 & 0.989 & 0.289 \\
   \hline
\end{tabular}
\caption{Total variation distance of predicted model to the true model from a
Gaussian mixture with 6 modes. Results are shown for varying $\alpha$ and grid
sizes. The total variation of the simple histogram (i.e., $\bar{y}$)
estimator is shown for comparison.}
\label{tvSim}
\end{table*}

\begin{table*}[t!]
\centering
\begin{tabular}{c|ccccccccc}
  \hline
num.  & \multicolumn{9}{|c}{$\alpha$ parameter} \\
tiles & 0 & 0.1 & 0.25 & 0.33 & 0.5 & 0.66 & 0.75 & 0.9 & 1 \\
  \hline
  $4^3$ & 1 & 1 & 3 & 3 & 2 & 2 & 4 & 2 & 1 \\
  $4^4$ & 33 & 15 & 15 & 15 & 24 & 48 & 48 & 3 & 3 \\
  $4^5$ & 59 & 63 & 76 & 105 & 128 & 128 & 128 & 2 & 1 \\
  $4^6$ & 63 & 87 & 144 & 161 & 198 & 174 & 182 & 198 & 0 \\
  $4^7$ & 66 & 87 & 145 & 167 & 199 & 167 & 168 & 171 & 6 \\
   \hline
\end{tabular}
\caption{Number of tiles with non-zero weights for the predicted model over
the simulated Gaussian mixture with 6 modes. Results are shown for varying $\alpha$ and grid
sizes, and use a constant $\delta$ equal to $0.001$.}
\label{sizeSim}
\end{table*}

\subsection{Fast query techniques}

Given the quadtree nature of dictionary $D$, any particular grid point $x_i$ is
represented by at most $k + 1 = \log_4 (n) + 1$ elements. This implies that we
can calculate the predicted value at that point in $\mathcal{O}(\log(n))$
time. Due to the nature of the $\ell_1$-penalty, the number of non-zero terms of
$\widehat{\beta}$ should be small; we denote this by $|| \widehat{\beta} ||_0 = s$.
Given the hardthresholding in our example, $s$ is at most $\delta^{-1}$ but will
typically be much smaller. From this it is possible to calculate
$\sum_{i\in A} x_i$ in $\mathcal{O}(s \log(n))$ time.

Assume now that we have two estimates density estimates, $\widehat{b}_A$ and $\widehat{b}_B$,
which represent the densities $\mathbb{P}(x_i | A)$ and $\mathbb{P}(x_i | B)$ for
disjoint events $A$ and $B$. In order to estimate $\mathbb{P}(x_i | A \cup B)$, we
can take the weighted sum of their sparse representations:
\begin{align}
\widehat{b}_{A \cup B} &= \mathbb{P} A \cdot \widehat{b}_A + \mathbb{P} B \cdot \widehat{b}_B
\end{align}
The final representation may be further hard thresholded by $\delta$ to guarantee
that $\widehat{b}_{A \cup B}$ has no more than $\delta^{-1}$ elements. We can
analogously use the same method to take the union of an large set of densities.

Finally, consider observing two independent samples in which we first observe event
$A$ and then observe the event $B$. Conditioned on the fact that both were observed
in the same location, we wish to calculate the density over space given the
estimators $\mathbb{P}(x_i | A)$ and $\mathbb{P}(x_i | A)$. We call the event of
observing $A$ followed by $B$ in the same location` $C$'.
Applications of this type of event arise when a single person or device is observed
displaying two types of events in close temporal proximity to one another (in other
words, too fast to have moved given the granularity of the grid).

We cannot directly calculate this density of the non-zero elements of the sparse
estimators, however clearly we have the following relationship between the sparse
representations:
\begin{align}
D^{(1)} \widehat{b}_C &= D^{(1)} \widehat{b}_A \cdot D^{(1)} \widehat{b}_B \label{prodCalc}
\end{align}
Calculating this directly would require computing estimates of the original densities
at all $n$ grid points. However, notice that because of the hierarchical structure
of the dictionary, $D^{(1)} \widehat{b}_A$ has at most $||\widehat{b}_A||_0 + 1$ unique values.
This is due to the fact that for any two tiles either one contains the other or
they are disjoint. Likewise, the product $D^{(1)} \widehat{b}_A \cdot D^{(1)} \widehat{b}_B$
has at most $||\widehat{b}_A||_0 + ||\widehat{b}_B||_0 + 1$ unique values. Therefore
the unique values of the right hand side of Equation~\ref{prodCalc} can be calculated
in $\mathcal{O}(s)$ time. Similarly $\widehat{b}_C$ can be solved for in $\mathcal{O}(s)$
time because it is known that the solution lies in the joint support of
$\widehat{b}_A$ and $\widehat{b}_B$.

\begin{figure*}[t!]
    \centering
    \includegraphics[width=\textwidth]{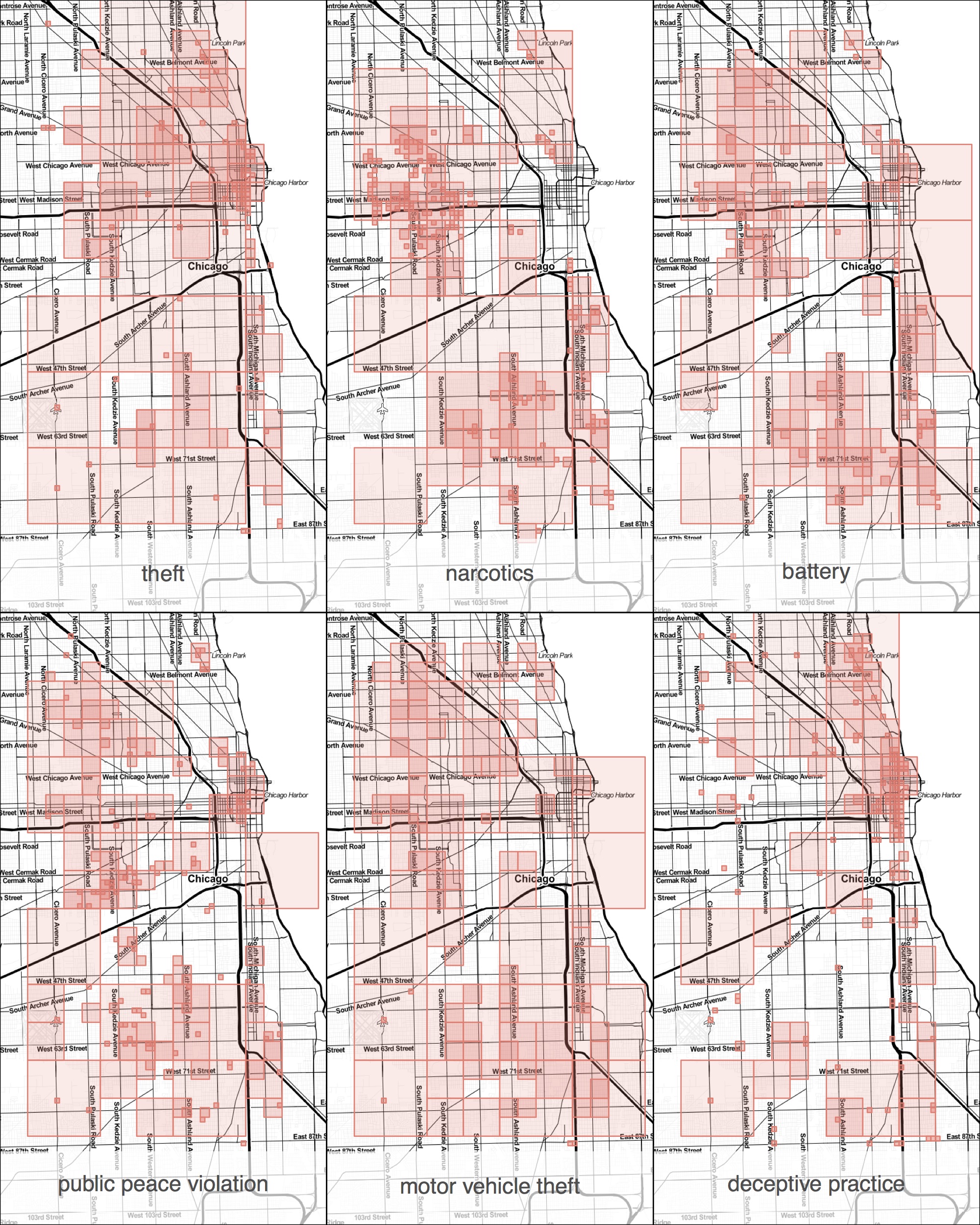}
    \caption{Six predicted models for the spatial density of the occurance
    of six times of crimes in the Chicago crime sataset. The $\alpha$ parameter
    is fixed to be $0.4$ across all models. Map tiles by Stamen Design.}
    \label{chiimage}
\end{figure*}

\begin{table*}[t!]
    \centering
    \begin{tabular}{@{} cl*{7}c @{}}
        \hline
            &&  &  &  & peace
            & vehicle  & deceptive  \\
            && theft & narcotics & battery &  violation
            &  theft &  practice \\
        \hline
          & theft & 143 & 226 & 210 & 195 & 154 & 171 \\
          & narcotics &  & 239 & 241 & 254 & 219 & 257 \\
          & battery &  &  & 156 & 226 & 195 & 244 \\
          & peace violation &  &  &  & 191 & 197 & 228 \\
          \rot{\rlap{~Union}}
          & vehicle theft &  &  &  &  & 107 & 187 \\
          & deceptive practice &  &  &  &  &  & 186 \\
        \hline
          & theft & 143 & 227 & 190 & 199 & 157 & 188 \\
          & narcotics &  & 231 & 266 & 287 & 246 & 151 \\
          & battery &  &  & 152 & 218 & 157 & 168 \\
          & peace violation &  &  &  & 174 & 197 & 166 \\
          & vehicle theft &  &  &  &  & 106 & 168 \\
          \rot{\rlap{~Intersection}}
          & deceptive practice &  &  &  &  &  & 173 \\
        \hline
    \end{tabular}
  \caption{Number of non-zero tile weights in the Chicago crime data,
  for pairwise unions and intersections with $\alpha=0.4$ and $\delta=0.001$.
  Note that self-unions return the original density.}
  \label{chiSize}
\end{table*}

\begin{table*}[t!]
\centering
\begin{tabular}{r|rrrrrrrrrrrr}
  \hline
  & \multicolumn{12}{|c}{start hour} \\
stop hour & 00 & 02 & 04 & 06 & 08 & 10 & 12 & 14 & 16 & 18 & 20 & 22 \\
  \hline
  02 & 189 & 174 & 184 & 189 & 188 & 179 & 171 & 175 & 168 & 166 & 161 & 163 \\
  04 &  & 196 & 187 & 199 & 191 & 182 & 170 & 167 & 165 & 163 & 164 & 162 \\
  06 &  &  & 208 & 190 & 188 & 188 & 175 & 166 & 164 & 161 & 158 & 158 \\
  08 &  &  &  & 230 & 199 & 185 & 171 & 167 & 160 & 157 & 160 & 161 \\
  10 &  &  &  &  & 210 & 166 & 160 & 160 & 158 & 157 & 157 & 156 \\
  12 &  &  &  &  &  & 188 & 162 & 160 & 163 & 165 & 158 & 153 \\
  14 &  &  &  &  &  &  & 183 & 161 & 170 & 160 & 154 & 158 \\
  16 &  &  &  &  &  &  &  & 187 & 173 & 160 & 160 & 162 \\
  18 &  &  &  &  &  &  &  &  & 194 & 159 & 163 & 164 \\
  20 &  &  &  &  &  &  &  &  &  & 175 & 159 & 166 \\
  22 &  &  &  &  &  &  &  &  &  &  & 190 & 170 \\
  24 &  &  &  &  &  &  &  &  &  &  &  & 197 \\
   \hline
\end{tabular}
    \caption{The number of non-zero coefficients attained by taking
    the union of contiguous ranges from the two-hour (local time) Uber pickup densities.
    All combinations of inter-day ranges (that is, not crossing midnight) between even hours
    are shown. All hours are in local time.}
    \label{unionSize}
\end{table*}

\begin{table*}[t!]
\centering
\begin{tabular}{r|rrrrrrrrrrrr}
  \hline
  & \multicolumn{12}{|c}{start hour} \\
stop hour & 00 & 02 & 04 & 06 & 08 & 10 & 12 & 14 & 16 & 18 & 20 & 22 \\
  \hline
  02 & 201 & 223 & 256 & 259 & 250 & 243 & 217 & 180 & 150 & 116 & 84 & 45 \\
  04 &  & 194 & 273 & 292 & 276 & 261 & 257 & 230 & 190 & 150 & 108 & 62 \\
  06 &  &  & 208 & 266 & 259 & 251 & 245 & 227 & 186 & 146 & 97 & 62 \\
  08 &  &  &  & 228 & 256 & 256 & 250 & 244 & 214 & 182 & 167 & 119 \\
  10 &  &  &  &  & 208 & 243 & 231 & 224 & 215 & 174 & 170 & 152 \\
  12 &  &  &  &  &  & 188 & 203 & 195 & 190 & 175 & 172 & 167 \\
  14 &  &  &  &  &  &  & 183 & 203 & 196 & 191 & 188 & 177 \\
  16 &  &  &  &  &  &  &  & 187 & 203 & 197 & 206 & 193 \\
  18 &  &  &  &  &  &  &  &  & 195 & 211 & 222 & 226 \\
  20 &  &  &  &  &  &  &  &  &  & 176 & 214 & 221 \\
  22 &  &  &  &  &  &  &  &  &  &  & 190 & 224 \\
  24 &  &  &  &  &  &  &  &  &  &  &  & 198 \\
   \hline
\end{tabular}
    \caption{The number of non-zero coefficients attained by taking
    the intersection of contiguous ranges from the two-hour (local time) Uber pickup densities.
    All combinations of inter-day ranges (that is, not crossing midnight) between even hours
    are shown. All hours are in local time.}
    \label{intersectSize}
\end{table*}

\begin{figure*}[t!]
    \centering
    \includegraphics[width=\textwidth]{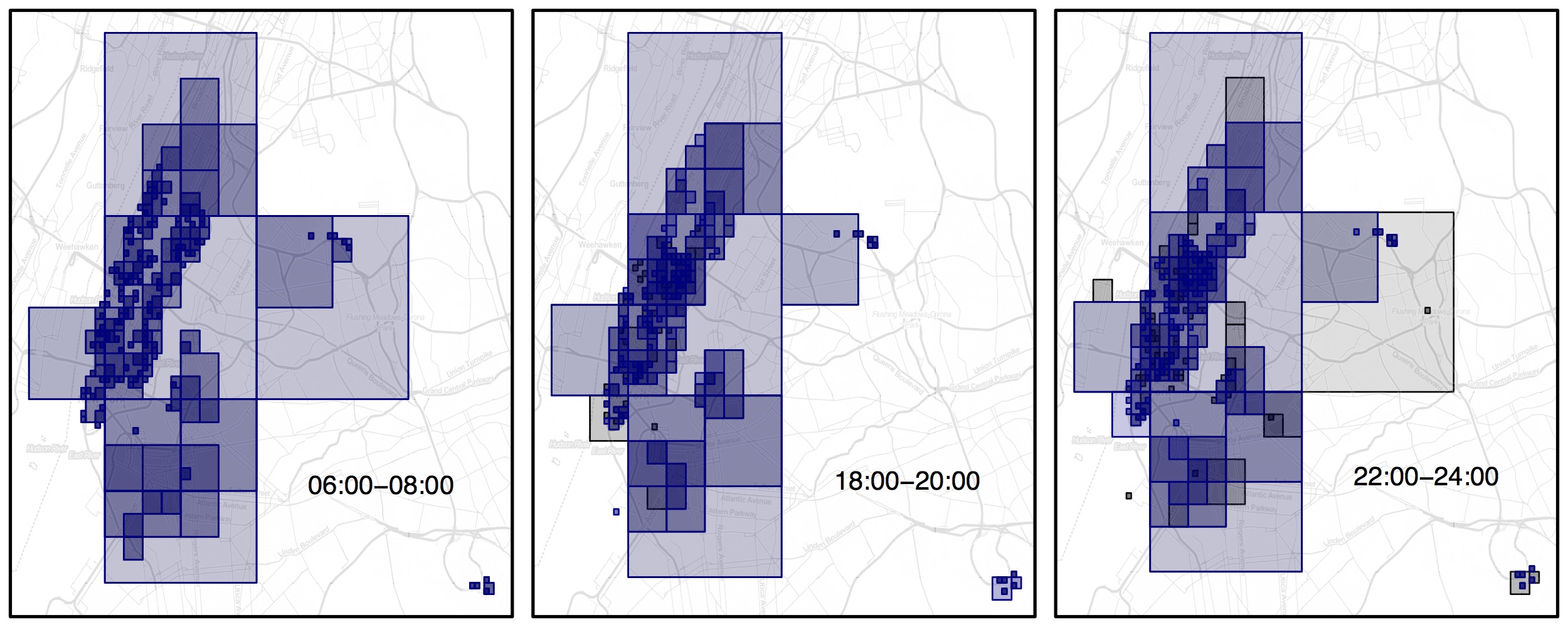}
    \caption{Three predicted models for the spatial density of the Uber
    pickup location dataset for two hour intervals. The $\alpha$ parameter
    is fixed to be $0.4$. All hours are in local time.
    Map tiles by Stamen Design.}
    \label{uberimage}
\end{figure*}

\section{Simulation from a Gaussian mixture model}

We sampled a set of two hundred thousand points from a Gaussian mixture model with
$6$ modes. Using $\delta = 0.001$, we calculated the sparse representation of
the density over the quadtree based dictionary. A scatterplot of the sampled
points  superimposed over the predicted densities are shown in
Figure~\ref{sim1image} for a particular grid size and value of $\alpha$.
Table~\ref{tvSim} displays the total variation distance between the estimated
density and the true Gaussian distribution over a grid of tile sizes and values
of $\alpha$. These are compared to the total variation distances gained by
directly using the observed densities at each point. We see that,
with the exception of the smallest grid size, the total
variation of both estimators performs more poorly with larger grid sizes.
This seems reasonable as larger grid sizes provide a harder estimation problem
with many more degrees of freedom.
The total variation as a function of $\alpha$ shows that values too
close to $1$ perform poorly regardless of the grid size. Values near $0$ function
okay, but the optimal value tends to be from $0.25$ to $0.75$.

In Table~\ref{sizeSim}, we show the number of non-zero elements in the prediction
vector by grid size and the $\alpha$ parameter. The number of non-zero elements
generally increases with the grid size. Given the exponentially growing dictionary
size this is not surprising, however it is interesting that the sizes do not grow
exponentially. This shows the ability of our method to represent complex, granular
densities in a relatively compact way regardless of the grid size. When $\alpha$
is equal to $1$ (or very close to it) the predicted model size is very small. This
is the result of a poor model fit due to overfitting on small tile sizes; the
cross-validation procedure eliminates the overfitting and leaves a relatively
constant model.
% Note that these correspond to bad total variation distance in
% Table~\ref{tvSim}. The model size is typically largest around $\alpha$ equal to
% $0.5$; smaller tuning parameters put most of the mass on the larger tiles and
% larger tuning parameters tend to overfit and become pruned by the
% cross-validation step.
We recommend setting $\alpha$ to anything between $0$
and the value that maximized the model size. In this simulation notice that
maximizing the model size with respect to $\alpha$ would generate the lowest
total variation of the resulting model. This is a useful heuristic because
while Table~\ref{tvSim} cannot be generated without knowing the true generating
distribution, Table~\ref{sizeSim} can be created without an external information.

\section{Applications}

\subsection{Chicago crime data}

The city of Chicago releases incident level data for all reported crimes
that occur within the city \cite{chicrimedata}. We fit sparse density
estimates to six classes of crimes; these estimated densities are shown
in Figure~\ref{chiimage}. We have picked classes that exhibit very
different spatial densities over the city. For example, deceptive practice
crimes occur predominantly in the center of the city whereas narcotics
violations are concentrated in the western and southern edges of Chicago.
Table~\ref{chiSize} shows the number of non-zero terms in the
predicted density vectors of pairwise unions and intersections. Of
particular importance, note that the complexity of the unions and
intersections are not significantly larger than the complexity of the
original estimates (given by diagonal terms of the table of unions).
This property would not hold for most other density estimation algorithms.

\subsection{Uber pickup locations in NYC}

In response to a freedom of information request, the New York City government
released a dataset showing the requested pickup locations from $5$ million
rides commissioned by the transit company Uber \cite{uber2015}.
We used this data to construct two-hour density buckets, three of which
are shown in Figure~\ref{uberimage}. Notice that,
unlike the Chicago crime data, the non-zero tiles are relatively consistent
from image to image, roughly following the population density of the city.
Temporal differences do exist: for example the heavily
neighborhood of the Upper West Side has a particularly high density only during
the morning commute.
% Williamsburg, a trendy neighborhood of bars and
% restaurants likewise has a far greater density between 22:00 and 24:00.
% There are also localized spikes at the airports JFK and LaGuardia, popular
% pick-up locations for taxi rides.
The ability to detect localized spikes at the airports
exhibit the adaptive nature of the sparse learning algorithm.

Table~\ref{unionSize} shows the model sizes when computing the
union of densities from any continuous time interval during the day.
Due to the truncating of small densities by $\delta$ and the fact that
the non-zero tiles generally line up across time periods, the overall
size of the unions never grows much larger than the original estimates.
Table~\ref{intersectSize} shows the same information over arbitrary unions.
These intersections would be useful, for example, when trying to determine
where taxicab waiting spots should be constructed as they indicate areas of
high density throughout periods of the day.
% The complexity of intersection also never grow too larger, with
% the longest time periods actually having significantly smaller tile sets
% as densities collapse over areas that are only frequented all day.

Overall, we are able to quickly calculate $144$ complex densities by only
estimating and storing $12$ of them (or $4095$, when considering non-contiguous
time periods).

\section{Conclusions and future extensions}

We have presented an algorithm for calculating sparse representations
of spatial densities. This method has been shown to be able to compute
fast density estimations over arbitrary regions and to support union
and intersections over a large set of independent density estimates.
These claims have been illustrated theoretically, by controlled simulations,
and over two real datasets.
% We are currently using this approach to simultaneously manipulate
% millions of density estimates that can be stored directly in memory
% on a single server.
% We are now looking to integrate the sparse
% tile densities directly into a standard spatial database system and
% to calculate set operations between densities within the database
% itself. This would allow for faster queries and would facilitate
% storing larger collections of densities on even smaller hardware.
We are now looking to generalize the quadtree approach to work with
alternative hierarchical partitions of space. In particular, this could
create a finer grained dictionary near places of high density (i.e.,
roads and city centers) allowing for a smaller dictionary and
further decreasing the estimation error.

\clearpage
\newpage

\subsubsection*{References}

\bibliographystyle{abbrvnat}

\begingroup
\renewcommand{\section}[2]{}

\endgroup

\end{document}